\documentstyle[12pt,aps,epsfig]{revtex}
\vspace{2cm}
\begin{document}

\bibliographystyle{prsty}

\title{\ \\ \ \\ \ \\ \ \\ \ \\  \ \\
Transport in an interacting wire connected to measuring leads and proximity effects}
\author{I. Safi, H. J. Schulz}
\address{Laboratoire de Physique des Solides, Universit\'e Paris-Sud, 91405
Orsay, France}

\maketitle
\thispagestyle{empty}
\bigskip

{\small We investigate transport through a finite interacting wire connected
to noninteracting leads. The conductance of the pure wire is not
renormalized by the interactions for any spatial variation of the
interaction parameters $u,K$, and not even for Coulomb interactions
restricted to the wire. We rigorously relate the conductance to the
transmission, that turns out to be perfect. If $K$ varies abruptly at the
contacts, an electron incident on the wire is reflected into a series of
partial spatially separated charges which sum up to unity. For attractive interactions, the reflection at the contact is similar to Andreev reflection on a gapless superconductor. This process
affects the density-density or pairing correlation functions: they are
enhanced on the bulk of the wire as in an infinite Luttinger liquid, then
extend to the external noninteracting leads in a way reminiscent of the
proximity effect. The effect of impurities is governed by the wire parameter
but is affected by the leads close to the contacts. Our results give a
possible explanation to recent experiments on quantum wires by Tarucha et
al. \cite{tarucha}.}
\bigskip

\section{Introduction}
Quantum wires open a new perspective to test the predictions of the
well--developed theory of electron--electron interactions in one dimension,
generically described as a Luttinger liquid. One of the theoretical
predictions \cite {apel_rice,kane_fisher} is the renormalization of the
conductance by the interactions, $g=Ke^2/h$, where $K$ is a key parameter
depending on the interactions, with $K=1$ for a noninteracting
system. Recently, Tarucha et al.  \cite{tarucha} studied relatively clean
and long quantum wires: at high enough temperatures, when the impurities do
not affect the transport, the conductance is quantized in units of
$e^2/h$. Upon lowering the temperature, the impurities become effective, and
the observed decrease in the conductance fits the power law behavior
predicted by the Luttinger theory.  Tarucha and al. extract a parameter
$K\simeq 0.7$, which contradicts $ g=e^2/h\neq 0.7e^2/h$.

The model we propose provides a possible explanation for this paradox
\cite{ines}\footnote{The model has been proposed simultaneously by D. Maslov
and M. Stone \cite{maslov_g}, whose contribution is included in this
volume}. In contrast to previous transport results that overlook boundary
effects in the wire, we take into account the measuring leads, supposed to be one-dimensional. The global system thus formed is treated as
a Luttinger liquid with inhomogeneous parameters, $u(x),K(x)$ with $K$ set
to unity and $u$ to the Fermi velocity $v_F$ on the external leads. The
Hamiltonian is:
\begin{equation}
\label{Hgen}
H=\int \frac{dx}{2\pi }\left[ \frac {u}{K}\left( \partial _x\Phi
\right) ^2+uK\left( \partial _x\Theta \right) ^2\right]
\end{equation} 
where the boson field $\Phi $ is related to the particle density by $\rho
-\rho _0=-\partial _x\Phi /\pi $, and $\partial _x\Theta /\pi $ is the
field canonically conjugate to $\Phi $. The system is now similar to an
elastic string with inhomogeneous sound velocity $u(x)$ and compressibility
$u(x)/K(x) $. We restrict ourselves to spinless electrons for simplicity.

\section{Conductance of the pure wire}

In order to compute the conductance, we simulate the external reservoirs by
the potential they impose on the asymptotic regions of the external leads.
In the stationary regime, the current through the system depends only on
these asymptotic values. There are many ways to compute the conductance, but
the most straightforward and physically appealing one is through its
relation to the transmission. There is no general Landauer formula for our
interacting wire, where even the description in terms of quasiparticles
fails; for the present model however we could establish it
\emph{rigorously}.\cite{ines}

Next we have to find the transmission of an incident electron on the
interacting wire. For this purpose, we derive the equation of motion for the
particle density, with the initial condition given by the injected charge of
the incident electron. The total transmission turns out to be perfect! Thus
the conductance is given by:
\begin{equation}  \label{eq:conductancepur}
g=e^2/h.
\end{equation}
It is worth noting that the perfect transmission allows to use the underlying argument of Landauer's formula, where the reservoirs are simulated by the flux they inject. Our proof applies to any variation of the interaction parameters on the
wire, but not only short-range interactions: it holds for any Coulomb
interactions perfectly screened on the measuring leads, such as
$$U(x,y)=\frac{f(x)f(y)}{\left| x-y\right|} $$ with $f$ vanishing outside a
finite region around the wire.\\

\section{Dynamic process of transmission}

\begin{figure}[htb]
\begin{center}
\mbox{\epsfig{file=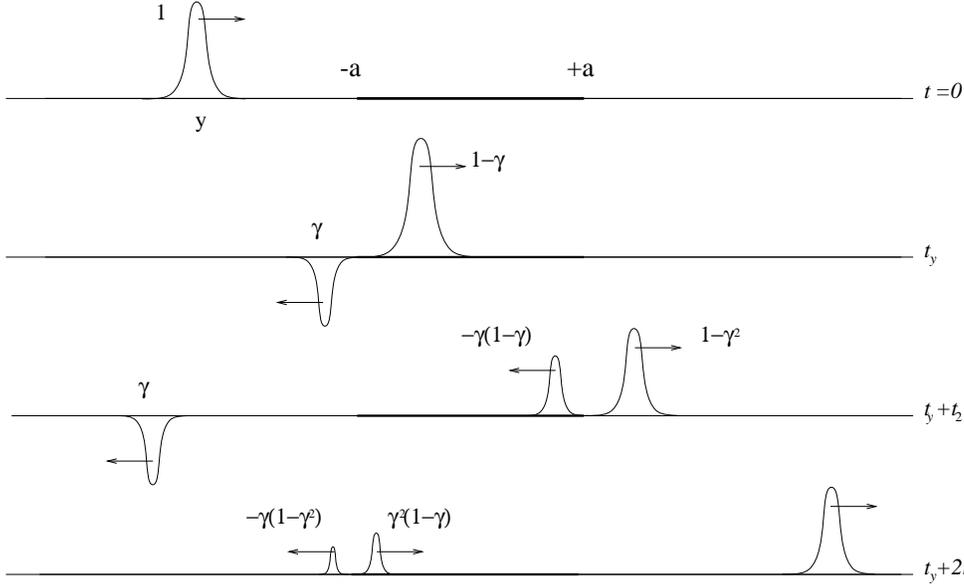,width=13 cm}}
\caption[to]{The transmission process of an incident electron on the wire in
the case where $K>1$ and $u=v_F$ for simplicity. We denote:
$t_y=(-a-y)/v_F$, and $t_2=2a/u=1/T_L$ the traversal time of the wire by the
density oscillations. At $t_y+(2n+1)t_2$ (resp. $t_y+2nt_2$), a charge
$\gamma^{2n}(1-\gamma^2)$(resp. $-\gamma^{2n-1}(1-\gamma^2)$) comes out at
$a$(resp.$-a$). The first reflected charge is of hole type, while the
subsequent ones are of electron type. If $K<1$, the signs of the reflected
charges are exchanged.\label{transmission}}
\end{center}
\end{figure}

It is not obvious to guess the perfect transmission when the interactions
are abruptly switched on the wire. Let's restrict ourselves to short-range
interactions, with a parameter $K$ on the wire, a situation we deal with in
the rest of the paper. An incident electron starts to interact with the
electrons of the wire, and is partially reflected at the contact with
coefficient
\footnote{This coefficient comes from two matching conditions of the boson
field at the contacts, required by the equation of motion: the continuity of
the current (the interactions conserve momentum) and of $(u/K)\rho$}
\[
\gamma =\frac{1-K}{1+K} 
\]
How can it be transmitted in the stationary regime? This is because the
subsequent reflections are of hole (electron) type if the first one is of
electron (hole) type, which is the case of repulsive (attractive)
interaction: this is illustrated in fig.\ref{transmission}. The incident
electron is transmitted into a series of spatially separated partial charges
that sum up to unity for times long compared to the traversal time of the
wire. The wire behaves as a Fabry-Perot resonator. 

\section{Proximity effect}

Before considering the effect of impurities on the conductance, let's ask
the following question: we know that an infinite Luttinger liquid has a
tendency towards superconducting order or the formation of a charge density
wave, depending on whether the interactions are attractive or
repulsive. This will make it a more or less good conductor. Now that we
connect it to measuring leads, the conductance is always equal to the
noninteracting value $e^2/h$. Does the tendency to one or the other type of
order persists?\ Yes, it does, and even more it extends in the
noninteracting leads, recalling the proximity effect.

The previous dynamic process due to change in interactions affects the
correlation functions: note that the latter depend on the temperature, while
the transport and transmission did not. Let us for instance focus on the
charge density correlation functions. In the high temperature limit, when
the distance between the contacts is higher than the thermal coherence
length $L_T=u/T$, the multiple reflections are washed out, so that we
recover the same behavior in the bulk as in an infinite Luttinger liquid
with parameter $K$.\ But there is still the first reflection on the contacts
with coefficient $\gamma $, that affects the charge density correlation in
their vicinity up to distances of the order of $L_T$: it reduces (enhances)
them whenever the interactions on the wire are repulsive (attractive).\ This
yields an effective local parameter $K_a=2K/(1+K)=1-\gamma $, proportional
to the transmission coefficient of an incident electron on the contact:\
$K_a $ is greater (smaller) than $K$ for repulsive (attractive)
interactions.\ The external noninteracting leads moderate the effect of
interactions near the contacts, but they also feel the effect of the
interactions within the wire: indeed, the charge density correlations are
enhanced (reduced) in the leads if $K<1$ ($K>1$), up to a distance given
by $L_T$.\

Now we go down in temperature, until $L_T\gg L$: there is thermal
coherence all over the wire, and the multiple reflections affect the
correlation function, whose dependence on the temperature is now determined
by the external leads. Since the latter are noninteracting, there is no
dependence on temperature.\ But there is a dependence on the wire length,
that a transmitted charge from the leads have crossed many times! For
points either very close to the contacts or far from them by a distance of
the order of $L$, i.e. well on the bulk, the local correlation function can
be obtained by the substitution $T\rightarrow T_L=u/L$ in the high
temperature result, thus is governed respectively by $K_a$ and $K$.\ For
other points, this substitution no more works. We summarize the results in a simplified form in the table \ref{t:effect}. The charge density correlations are still
enhanced (reduced) compared to their noninteracting value (reached far away
in the external leads) if $K<1$ ($K>1$): this enhancement (reduction)
extends in the external leads up to a distance $\sim L$.

The same reasoning holds for the pairing correlation functions: they are
enhanced (reduced) whenever the charge density correlation are reduced
(enhanced), so that it is easy to translate the previous statements.\ The
local parameter at the contact is obtained by taking the inverse of $K$ in $%
K_a$, thus it is the average of $K$ with the lead parameter: $(1+K)/2$.
Whenever the interactions are attractive, $K>1$, the tendency towards
superconducting order extends in the noninteracting leads up to the shorter
length scale.\ For instance, at an external frequency $\omega \ll T_L$, and
at $T\ll T_L,$ the longer is the interacting wire, the most enhanced and
the most extended in the leads are the pairing correlations.
Now let's come back to the reflection of an incident electron on the
contact, that affected the correlation functions as already explained: when 
$K>1$, we have $\gamma <0$, thus a partial hole is reflected back: this is
the analogous of an Andreev reflection at the interface between a normal
metal and a gapless superconductor: the electron energy is obviously greater
than the vanishing gap, thus the reflection is partial. In our case, we get
exactly one hole reflected in the limit $K\gg 1$.\footnote{In Ref.\onlinecite{ines}, we defined a local conductance at the contact, given by $g_+=g_0K_a$. It verifies $g_0K_1\leq g_+\leq 2 g_0K_1$ recalling similar inequalities at a N-S interface}\\

\begin{table}[bht]
$$
\begin{array}{|c|c|}
\hline
\omega<T_L & T_L^{K_a-1}T_x^{K-K_a}\\
\hline
T_L<\omega<T_x &\omega^{K_a-1}T_x^{K-K_a}\\
\hline
T_x<\omega & \omega^{K-1}\\
\hline
\end{array}
$$
\caption{The local CDW correlation function for different points $x$ on
the interacting wire,
divided by its value in a noninteracting wire. All the energies have to be
divided by the bandwidth $\Lambda$ . $T_x=u/(a-|x|)$ is the
inverse of the time taken by a plasmon to go from $x$ to the closest contact.
 The behavior on the external lead can be deduced simply by
replacing $K$ by unity, and $T_x$ by $v_F/(|x|-a)$. $\omega$ is an energy
variable that can be either a frequency, the inverse of time, or the
temperature. When  $T_x\sim
T_L$ (resp. $T_x\sim
\Lambda$), the
behavior is governed only by $K$ ($K_a=2K/(1+K)$). The pairing
correlation function can be inferred by replacing $K$ by $1/K$, thus $K_a$
by $\overline{K}_a=2/(1+K)$. On the
external leads, we let $K\rightarrow 1$, but keep $\overline{K}_a$.}
\label{t:effect}
\end{table}

\section{Backscattering potential}

The conductance of the pure wire did not get affected by the reflections at
the contacts due to change in interactions: the interactions as well as
those reflections conserve the total momentum of the electrons. Obviously,
this is no more the case in the presence of a backscattering potential. Does
the Luttinger liquid have a chance to show up in the dirty wire? The effect
of backscattering is determined by the charge density correlations:
since we've already seen their sensitivity to interactions, we can already
guess the answer: yes, the interactions will affect the conductance when
electrons are backscattered.\ We refer to the previous discussion of the
inhomogeneity of the charge density correlation function, showing how the
reduction in the conductance depends on the impurity location. The table
\ref{t:effect} gives the correction to the conductance in the presence of a
barrier at $x$, by letting $\omega=T$.

We performed a renormalization procedure explicitely at finite temperature
for a barrier placed in our finite wire \cite{ines_nato}: we verify explicitely that the
interactions are not renormalized by the barrier, so that the
perturbative computation of the conductance is sufficient .

Let's make more comments about the barrier at the contact: this is indeed a
simple way of modeling a mismatch at the opening of a quantum wire into the
two-dimensional gas. The interactions are expected to be repulsive, thus $%
K_a<1$, and the conductance can be notably decreased in the limit of a long
wire. But when we get long wires, we can't avoid impurities in their bulk:
according to our results, the latter dominate the scattering at the contacts
because $K<K_a<1$.\ For the contact to dominate, the interactions have to be
attractive, $K>1$ in which case $1<K_a<K$: in this case, the external leads
reduce the local attraction, thus enhance the effect of the backscattering
near the contact.

Let's now consider an extended disorder on the wire, with Gaussian
distribution. There are many powers that emerge in the conductance.\ Let's
write the dominant ones, without giving the explicit coefficients:
\[
g=\frac{e^2}h\left[ 1-\frac L{l_e}\omega ^{2(K-1)}-\frac \alpha {l_e}\omega
^{2(K_a-1)}\right] 
\]
with $\omega =\max (T,T_L)/\Lambda $, $\Lambda $ being the bandwidth, and $%
\alpha \sim u/\Lambda $. The contribution from the impurities near the
contacts dominates for $K>K_c=(3+\sqrt{1}7)/4$, at $T<T_L^{}$. \ 

If we restore the spin of electrons, the pure wire conductance is just
multiplied by $2$ since the transport depends only on the charge degrees of
freedom. Conceptually, the extension of the backscattering effects is easy,
but a more richer behavior emerges.\ This will be the subject of a separate
publication. \ 

We note finally the coherence of our results with the experiment by Tarucha
and al \cite{tarucha}. The ballistic conductance equals $2e^2/h$, and a
decrease with temperature with a power law saturating at $T<T_L$ is
observed. But the exponent in this power law yields the wire parameter only
if there is at least one impurity in the bulk of the wire. \ We cannot
really decide about the impurity distribution in the measured wires.\ If the
bulk of a wire is clean, the reduction comes exclusively from the
backscattering at the contacts, which would yield a parameter $K=0.5$,
different from the value $K=0.7$ one infer in the presence of impurities on
the bulk. The observed decrease of the conductance with the wire length is
not a strong argument for an extended disorder: a barrier yields also a
correction saturating at $L^{K(x)-1}$, with $K(x)=K$ on the bulk and $K_a$
near the contacts.

\section{Summary}
To summarize, the conductance of an interacting one--dimensional wire
connected to perfect one--dimensional measuring leads is equal to $2e^2/h$,
for any range of the interactions on the wire, and as long as they conserve
the momentum of the electrons. But the Luttinger liquid behavior gets
revealed in the presence of backscattering potential whose effect depends in
a a nontrivial way on its distribution through the wire.
\section{Acknowledgments}
I.S. would like to thank M. B\"uttiker, S. Datta, T. Giamarchi,
C. Gl\"attli, A. Gogolin, D.L. Maslov, C. Pasquier, and V. Wees for fruitful
discussions. We would like to thank the organizers of the XXXI Rencontres
des Morionds conference on ``Correlated Fermions and Transport in Mesoscopic
Systems'' for their invitation, as their efforts to make this nice meeting
holding. 

\medskip

\noindent

\vspace{2cm}
\begin{center}
\small{\bf Transport dans un fil quantique connect\'e \`a des fils de mesure} 
\end{center}
\bigskip

{\small 
Lorsqu'on connecte un liquide de Luttinger \`a des fils de mesure
parfaits, nous montrons que sa conductance n'est plus renormalis\'ee par les
interactions, contrairement au r\'esultat jusque l\`a admis. 
 Nous d\'emontrons ce r\'esultat
 non seulement pour un profil arbitraire des param\`etres sur le fil
central, mais aussi en pr\'esence d'interactions Coulombiennes \'ecrant\'ees sur les
fils de mesure.\\
Le moyen le plus direct de le montrer est d'\'etablir
rigoureusement que la conductance est donn\'ee par le coefficient de
transmission d'un \'electron \`a travers le fil, et que la transmission est
parfaite dans le r\'egime stationnaire. Ainsi, pour des interactions \`a courte port\'ee  branch\'ees
discontinuement sur le fil, un \'electron incident est transmis en une s\'erie de charges partielles
spatialement s\'epar\'ees, dont la somme vaut l'unit\'e.\\
 Si les interactions
sont attractives, la r\'eflexion d'un \'electron sur le contact est
l'analogue d'une r\'eflexion d'Andreev. Les fonctions de corr\'elation
correspondant \`a la tendance \`a l'ordre 
supraconducteur sont renforc\'ees en s'\'etendant dans les fils externes,
rappelant ainsi l'effet de proximit\'e.\\
En pr\'esence d'impuret\'es, la conductance devient sensible aux
interactions, mais d\'epend \`a la fois de l'emplacement des impuret\'es et
des param\`etres du fil mesur\'e ainsi que celui des fils de mesure. Nos r\'esultats donnent une explication
possible au paradoxe soulev\'e par les exp\'eriences r\'ecentes de Tarucha
et al sur des fils quantiques.
}
\bigskip

\end{document}